\documentclass[twocolumn,letterpaper,aps,prl,superscriptaddress,showpacs,floatfix]{revtex4}

\usepackage{graphicx}	
\usepackage{xspace}     

\newcommand{\raa}{\mbox{$R_{\rm AA}$}\xspace}

\newcommand{\sqs}{\mbox{$\sqrt{s}$}\xspace}
\newcommand{\sqsn}{\mbox{$\sqrt{s_{_{NN}}}$}\xspace}

\begin{document}


\title{Direct photon production in $d$+Au collisions at 
$\sqrt{s_{_{NN}}}$=200~GeV}

\newcommand{\abilene}{Abilene Christian University, Abilene, Texas 79699, USA}
\newcommand{\acadsin}{Institute of Physics, Academia Sinica, Taipei 11529, Taiwan}
\newcommand{\banaras}{Department of Physics, Banaras Hindu University, Varanasi 221005, India}
\newcommand{\barc}{Bhabha Atomic Research Centre, Bombay 400 085, India}
\newcommand{\bnlcoll}{Collider-Accelerator Department, Brookhaven National Laboratory, Upton, New York 11973-5000, USA}
\newcommand{\bnlphys}{Physics Department, Brookhaven National Laboratory, Upton, New York 11973-5000, USA}
\newcommand{\caucr}{University of California - Riverside, Riverside, California 92521, USA}
\newcommand{\charlesczech}{Charles University, Ovocn\'{y} trh 5, Praha 1, 116 36, Prague, Czech Republic}
\newcommand{\chonbuk}{Chonbuk National University, Jeonju, 561-756, Korea}
\newcommand{\ciae}{Science and Technology on Nuclear Data Laboratory, China Institute of Atomic Energy, Beijing 102413, P.~R.~China}
\newcommand{\cns}{Center for Nuclear Study, Graduate School of Science, University of Tokyo, 7-3-1 Hongo, Bunkyo, Tokyo 113-0033, Japan}
\newcommand{\colorado}{University of Colorado, Boulder, Colorado 80309, USA}
\newcommand{\columbia}{Columbia University, New York, New York 10027 and Nevis Laboratories, Irvington, New York 10533, USA}
\newcommand{\czechtech}{Czech Technical University, Zikova 4, 166 36 Prague 6, Czech Republic}
\newcommand{\dapnia}{Dapnia, CEA Saclay, F-91191, Gif-sur-Yvette, France}
\newcommand{\debrecen}{Debrecen University, H-4010 Debrecen, Egyetem t{\'e}r 1, Hungary}
\newcommand{\elte}{ELTE, E{\"o}tv{\"o}s Lor{\'a}nd University, H - 1117 Budapest, P{\'a}zm{\'a}ny P. s. 1/A, Hungary}
\newcommand{\ewha}{Ewha Womans University, Seoul 120-750, Korea}
\newcommand{\fit}{Florida Institute of Technology, Melbourne, Florida 32901, USA}
\newcommand{\fsu}{Florida State University, Tallahassee, Florida 32306, USA}
\newcommand{\gsu}{Georgia State University, Atlanta, Georgia 30303, USA}
\newcommand{\hiroshima}{Hiroshima University, Kagamiyama, Higashi-Hiroshima 739-8526, Japan}
\newcommand{\ihepprot}{IHEP Protvino, State Research Center of Russian Federation, Institute for High Energy Physics, Protvino, 142281, Russia}
\newcommand{\illuiuc}{University of Illinois at Urbana-Champaign, Urbana, Illinois 61801, USA}
\newcommand{\inrras}{Institute for Nuclear Research of the Russian Academy of Sciences, prospekt 60-letiya Oktyabrya 7a, Moscow 117312, Russia}
\newcommand{\instpasczech}{Institute of Physics, Academy of Sciences of the Czech Republic, Na Slovance 2, 182 21 Prague 8, Czech Republic}
\newcommand{\isu}{Iowa State University, Ames, Iowa 50011, USA}
\newcommand{\jinrdubna}{Joint Institute for Nuclear Research, 141980 Dubna, Moscow Region, Russia}
\newcommand{\jyvaskyla}{Helsinki Institute of Physics and University of Jyv{\"a}skyl{\"a}, P.O.Box 35, FI-40014 Jyv{\"a}skyl{\"a}, Finland}
\newcommand{\kek}{KEK, High Energy Accelerator Research Organization, Tsukuba, Ibaraki 305-0801, Japan}
\newcommand{\korea}{Korea University, Seoul, 136-701, Korea}
\newcommand{\kurchatov}{Russian Research Center ``Kurchatov Institute", Moscow, 123098 Russia}
\newcommand{\kyoto}{Kyoto University, Kyoto 606-8502, Japan}
\newcommand{\labllr}{Laboratoire Leprince-Ringuet, Ecole Polytechnique, CNRS-IN2P3, Route de Saclay, F-91128, Palaiseau, France}
\newcommand{\lawllnl}{Lawrence Livermore National Laboratory, Livermore, California 94550, USA}
\newcommand{\losalamos}{Los Alamos National Laboratory, Los Alamos, New Mexico 87545, USA}
\newcommand{\lpc}{LPC, Universit{\'e} Blaise Pascal, CNRS-IN2P3, Clermont-Fd, 63177 Aubiere Cedex, France}
\newcommand{\lund}{Department of Physics, Lund University, Box 118, SE-221 00 Lund, Sweden}
\newcommand{\maryland}{University of Maryland, College Park, Maryland 20742, USA}
\newcommand{\mass}{Department of Physics, University of Massachusetts, Amherst, Massachusetts 01003-9337, USA }
\newcommand{\muenster}{Institut fur Kernphysik, University of Muenster, D-48149 Muenster, Germany}
\newcommand{\muhlenberg}{Muhlenberg College, Allentown, Pennsylvania 18104-5586, USA}
\newcommand{\myongji}{Myongji University, Yongin, Kyonggido 449-728, Korea}
\newcommand{\nagasaki}{Nagasaki Institute of Applied Science, Nagasaki-shi, Nagasaki 851-0193, Japan}
\newcommand{\newmex}{University of New Mexico, Albuquerque, New Mexico 87131, USA }
\newcommand{\nmsu}{New Mexico State University, Las Cruces, New Mexico 88003, USA}
\newcommand{\ornl}{Oak Ridge National Laboratory, Oak Ridge, Tennessee 37831, USA}
\newcommand{\orsay}{IPN-Orsay, Universite Paris Sud, CNRS-IN2P3, BP1, F-91406, Orsay, France}
\newcommand{\peking}{Peking University, Beijing 100871, P.~R.~China}
\newcommand{\pnpi}{PNPI, Petersburg Nuclear Physics Institute, Gatchina, Leningrad region, 188300, Russia}
\newcommand{\riken}{RIKEN Nishina Center for Accelerator-Based Science, Wako, Saitama 351-0198, Japan}
\newcommand{\rikjrbrc}{RIKEN BNL Research Center, Brookhaven National Laboratory, Upton, New York 11973-5000, USA}
\newcommand{\rikkyo}{Physics Department, Rikkyo University, 3-34-1 Nishi-Ikebukuro, Toshima, Tokyo 171-8501, Japan}
\newcommand{\saispbstu}{Saint Petersburg State Polytechnic University, St. Petersburg, 195251 Russia}
\newcommand{\saopaulo}{Universidade de S{\~a}o Paulo, Instituto de F\'{\i}sica, Caixa Postal 66318, S{\~a}o Paulo CEP05315-970, Brazil}
\newcommand{\seoulnat}{Seoul National University, Seoul, Korea}
\newcommand{\stonybrkc}{Chemistry Department, Stony Brook University, SUNY, Stony Brook, New York 11794-3400, USA}
\newcommand{\stonycrkp}{Department of Physics and Astronomy, Stony Brook University, SUNY, Stony Brook, New York 11794-3400, USA}
\newcommand{\subatech}{SUBATECH (Ecole des Mines de Nantes, CNRS-IN2P3, Universit{\'e} de Nantes) BP 20722 - 44307, Nantes, France}
\newcommand{\tenn}{University of Tennessee, Knoxville, Tennessee 37996, USA}
\newcommand{\titech}{Department of Physics, Tokyo Institute of Technology, Oh-okayama, Meguro, Tokyo 152-8551, Japan}
\newcommand{\tsukuba}{Institute of Physics, University of Tsukuba, Tsukuba, Ibaraki 305, Japan}
\newcommand{\vandy}{Vanderbilt University, Nashville, Tennessee 37235, USA}
\newcommand{\waseda}{Waseda University, Advanced Research Institute for Science and Engineering, 17 Kikui-cho, Shinjuku-ku, Tokyo 162-0044, Japan}
\newcommand{\weizmann}{Weizmann Institute, Rehovot 76100, Israel}
\newcommand{\wigner}{Institute for Particle and Nuclear Physics, Wigner Research Centre for Physics, Hungarian Academy of Sciences (Wigner RCP, RMKI) H-1525 Budapest 114, POBox 49, Budapest, Hungary}
\newcommand{\yonsei}{Yonsei University, IPAP, Seoul 120-749, Korea}
\affiliation{\abilene}
\affiliation{\acadsin}
\affiliation{\banaras}
\affiliation{\barc}
\affiliation{\bnlcoll}
\affiliation{\bnlphys}
\affiliation{\caucr}
\affiliation{\charlesczech}
\affiliation{\chonbuk}
\affiliation{\ciae}
\affiliation{\cns}
\affiliation{\colorado}
\affiliation{\columbia}
\affiliation{\czechtech}
\affiliation{\dapnia}
\affiliation{\debrecen}
\affiliation{\elte}
\affiliation{\ewha}
\affiliation{\fit}
\affiliation{\fsu}
\affiliation{\gsu}
\affiliation{\hiroshima}
\affiliation{\ihepprot}
\affiliation{\illuiuc}
\affiliation{\inrras}
\affiliation{\instpasczech}
\affiliation{\isu}
\affiliation{\jinrdubna}
\affiliation{\jyvaskyla}
\affiliation{\kek}
\affiliation{\korea}
\affiliation{\kurchatov}
\affiliation{\kyoto}
\affiliation{\labllr}
\affiliation{\lawllnl}
\affiliation{\losalamos}
\affiliation{\lpc}
\affiliation{\lund}
\affiliation{\maryland}
\affiliation{\mass}
\affiliation{\muenster}
\affiliation{\muhlenberg}
\affiliation{\myongji}
\affiliation{\nagasaki}
\affiliation{\newmex}
\affiliation{\nmsu}
\affiliation{\ornl}
\affiliation{\orsay}
\affiliation{\peking}
\affiliation{\pnpi}
\affiliation{\riken}
\affiliation{\rikjrbrc}
\affiliation{\rikkyo}
\affiliation{\saispbstu}
\affiliation{\saopaulo}
\affiliation{\seoulnat}
\affiliation{\stonybrkc}
\affiliation{\stonycrkp}
\affiliation{\subatech}
\affiliation{\tenn}
\affiliation{\titech}
\affiliation{\tsukuba}
\affiliation{\vandy}
\affiliation{\waseda}
\affiliation{\weizmann}
\affiliation{\wigner}
\affiliation{\yonsei}
\author{A.~Adare} \affiliation{\colorado}
\author{S.S.~Adler} \affiliation{\bnlphys}
\author{S.~Afanasiev} \affiliation{\jinrdubna}
\author{C.~Aidala} \affiliation{\columbia} \affiliation{\mass}
\author{N.N.~Ajitanand} \affiliation{\stonybrkc}
\author{Y.~Akiba} \affiliation{\kek} \affiliation{\riken} \affiliation{\rikjrbrc}
\author{H.~Al-Bataineh} \affiliation{\nmsu}
\author{A.~Al-Jamel} \affiliation{\nmsu}
\author{J.~Alexander} \affiliation{\stonybrkc}
\author{A.~Angerami} \affiliation{\columbia}
% ...
\collaboration{PHENIX Collaboration} \noaffiliation

\date{\today}


\begin{abstract}

Direct photons have been measured in $\sqrt{s_{_{NN}}}=200$~GeV $d$+Au 
collisions at midrapidity. A wide $p_T$ range is covered by measurements 
of nearly-real virtual photons ($1<p_T<6~$GeV/$c$) and real photons 
($5<p_T<16~$GeV/$c$). The invariant yield of the direct photons in $d$+Au 
collisions over the scaled $p$+$p$ cross section is consistent with unity. 
Theoretical calculations assuming standard cold-nuclear-matter effects 
describe the data well for the entire $p_T$ range. This indicates that the 
large enhancement of direct photons observed in Au$+$Au collisions for 
$1.0<p_T<2.5~$GeV/$c$ is due to a source other than the initial-state 
nuclear effects.

\end{abstract}

\pacs{25.75.Dw}
	
\maketitle



Direct photons in both Au$+$Au and $p$+$p$ collisions were measured at the 
Relativistic Heavy Ion 
Collider~\cite{ppg086,ppg042,ppg060,ppg136AXV,ppg139AXV} over a wide $p_T$ 
range, which was achieved through measurements of both real photons and 
nearly-real virtual photons~\cite{ppg088}.  For $1.0<p_T<2.5~$GeV/$c$, a 
significant excess of direct photons over the binary-scaled $p$+$p$ yield 
was observed in central Au$+$Au collisions, suggesting the existence of 
thermal photons emitted from the hot medium.  The key to measurements of 
the direct photon production for $p_T<5~$GeV/$c$ is the use of virtual 
photons, which greatly reduces the background of photons from 
$\pi^0,\eta\rightarrow2\gamma$.  For $p_T>4~$GeV/$c$, real photons are 
used and previous Au$+$Au measurements~\cite{ppg139AXV} indicate agreement 
with the binary-scaled $p$+$p$ collisions over $4<p_T<22~$GeV/$c$.  
However, effects either in the initial state or in the medium created in 
Au$+$Au collisions may cancel, making the $d$+Au measurement crucial to 
understanding the Au$+$Au results, because only initial-state effects are 
present in $d$+Au collisions.

Cold-nuclear-matter (CNM) effects may play an important role in direct 
photon production in A$+$A collisions and possibly modify the production 
rate compared to $p$+$p$ collisions.  CNM effects in the measured direct 
photon yield include interplays of various initial-state effects such as 
the Cronin enhancement~\cite{Cronin}, isospin effect, modification of the 
nuclear parton distribution functions (nPDFs) inside the 
nucleus~\cite{EMC,EPS09}, and the initial-state energy loss of colliding 
partons~\cite{ISE1,ISE2}.  The $d$+Au results shed light on these 
nontrivial effects and are necessary to make a firm statement about 
thermal photon emission in Au$+$Au collisions.  The CNM effects were 
studied in $d$+Au collisions at these energies through measurements of 
$\pi^0,\eta$ and $J/\psi$~\cite{ppg030,ppg044,ppg125AXV}; however, direct 
photons allow studying the initial-state nuclear effects -- without the 
ambiguities of the hadronization process.


In this paper, we present results of direct-photon measurements in 
\sqsn=200~GeV $d$+Au collisions at midrapidity for $1<p_T<16~$GeV/$c$.  
Both virtual-photon and real-photon measurements are performed as 
independent analyses.  The virtual-photon analysis uses data taken in 2008 
to provide results for the low $p_T$ region, approximately 
$1<p_T<6~$GeV/$c$.  The real-photon analysis uses data recorded in 2003 
for complimentary results above 5~GeV/$c$.  In addition, we report 
improved direct photon results in \sqs=200~GeV $p$+$p$ collisions for 
$1<p_T<5~$GeV/$c$ using 2006 data.  The new $p$+$p$ results are combined 
with the previously published $p$+$p$ collision data~\cite{ppg086,ppg088} 
from 2005 to serve as a reference for the $d$+Au data.

The two central arms of the PHENIX detector~\cite{PHENIX-NIM} cover 
$|\eta|<0.35$ in pseudorapidity and $\pi/2$ in azimuthal angle for each 
arm.  Minimum bias (MB) events were triggered by beam-beam counters 
located at both sides of the interaction point, covering $3.0<|\eta|<3.9$, 
which were also used to determine the event centrality for $d$+Au 
collisions.  Events containing high $p_T$ photons and electrons were 
selectively recorded by photon and single electron triggers in coincidence 
with the MB trigger. The photon trigger required an energy deposition in 
the electromagnetic calorimeter (EMCal) and the electron trigger required 
a hit in the ring imaging \v{C}erenkov detector with a correlated, above 
threshold, EMCal energy deposition. The virtual-photon analysis used 
0.7~nb$^{-1}$ of MB data and 54.9~nb$^{-1}$ of single-electron-triggered 
data.  The analyzed MB and single-photon-triggered data samples for the 
real-photon analysis were 0.8 and 1.6~nb$^{-1}$, respectively, where 
1~nb$^{-1}$ of $d$+Au collisions corresponds to $2\times197~$nb$^{-1}$ of 
nucleon-nucleon collisions.  We also analyzed 4.0~pb$^{-1}$ of the $p$+$p$ 
data from the 2006 run to measure the direct photon cross section for 
$1<p_T<5~$GeV/$c$ through the virtual photon analysis.


Electron tracks above 0.2~GeV/$c$ momentum are reconstructed using drift 
and pad chambers in each of the central arms, with momentum resolution 
$\sigma_{p_T}/p_T = 1.1\% \oplus 1.16\% \times p_T$.  Electrons are 
identified by requiring hits in the ring imaging \v{C}erenkov detector and 
matching the momentum with the energy measured in the EMCal.  Electron 
pairs are used to measure virtual photons using the method described in 
Ref.~\cite{ppg086,ppg088}.

Any source of real direct photons also produces nearly-real virtual 
photons, $i.e.$ low mass $e^+e^-$ pairs, allowing extraction of the real 
direct photon yield from low mass $e^+e^-$ pairs. In the virtual photon 
analysis, $e^+e^-$ pairs with $m_{ee}<0.3~$GeV/$c^2$ and pair 
$p_T>1~$GeV/$c$ are measured by the two central arms. Electron pairs are 
formed from combinations of all electrons and positrons with 
$p_T>0.3~$GeV/$c$ in an event, and background pairs arising from random 
combinations, external conversions, correlated background from double 
Dalitz decays of $\pi^0, \eta$ and jet induced correlations are removed by 
analysis techniques as discussed in Ref.~\cite{ppg088}. Electron pair mass 
distributions for different pair $p_T$ ranges, which comprise the virtual 
direct photon signal and the hadron decay component, are obtained. The 
inclusive photon yield is determined from the yield of $e^+e^-$ pairs in 
$m_{ee}\sim0.05~$GeV/$c^2$ with the relation of 
$\frac{d^2n_{ee}}{dm_{ee}}=\frac{2\alpha}{3\pi} 
\frac{1}{m_{ee}}dn_{\gamma}$~\cite{ppg088}. The $e^+e^-$ mass distribution 
for $m_{ee}<0.3~$GeV/$c^2$ and $p_T>1~$GeV/$c$ is decomposed by a 
two-component fitting procedure described in Ref.~\cite{ppg088} using the 
known shapes of the direct photon and hadron decay components. The direct 
photon fraction, $r_{\gamma}=$ direct $\gamma$/inclusive $\gamma$, is 
extracted from the fitting. Multiplying the direct photon fraction by the 
inclusive photon yield leads to the direct photon yield.

The systematic uncertainties on the direct-photon fraction are estimated 
from the difference in extracted direct-photon fraction when varying: (1) 
the particle compositions in the ``cocktail'' of hadron decay 
contributions for the fit, (2) the background subtraction of the measured 
mass distribution, (3) the mass region used for the fit, and (4) the 
efficiency corrections. The largest uncertainty is due to the particle 
composition of the hadronic cocktail, particularly 
$\eta/\pi^0=0.48\pm0.03$ at $p_T>2~$GeV/$c$, which is essentially 
identical to $p$+$p$~\cite{ppg051}. The resulting uncertainty in the 
direct-photon fraction due to $\eta/\pi^0$ is about 20--30\%, and less 
than 5\% are from all other sources. The uncertainty in the $e^+e^-$ pair 
acceptance correction introduces an additional 9\% uncertainty to the 
inclusive photon yield, which is added in quadrature with the other 
uncertainties.

\begin{figure}[tb]
\includegraphics[width=1.0\linewidth]{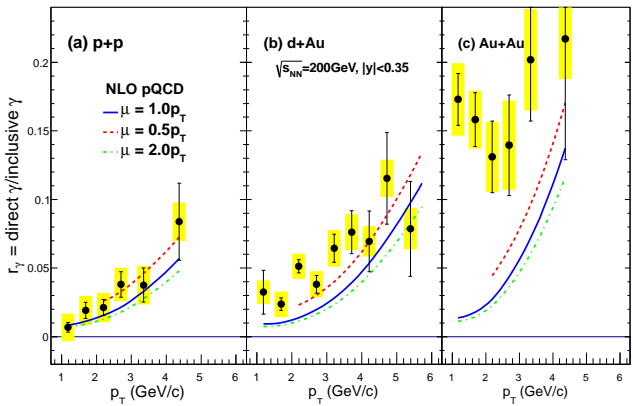}
\caption{\label{fig:r_gamma} (color online) 
The direct-photon fractions from the virtual-photon analysis as a function 
of $p_T$ in (a) $p$+$p$, (b) $d$+Au, and (c) Au$+$Au (MB)~\cite{ppg086} 
collisions. The statistical and systematic uncertainties are shown by the 
bars and bands, respectively.  The curves show expectations from a NLO 
pQCD calculation~\protect\cite{NLOpQCD,Werner} with different cutoff mass 
scales: (solid) $\mu=0.5~p_T$, (dash) $\mu=1.0~p_T$, and (dash-dot) 
$\mu=2.0~p_T$.
}
\end{figure}

Figure~\ref{fig:r_gamma} shows the measured direct-photon fractions by the 
virtual-photon analysis in $p$+$p$, $d$+Au, Au$+$Au~\cite{ppg086} 
collisions from left to right. The $p$+$p$ result is the combination of 
\cite{ppg086} and the 2006 data. The curves show the expectations from a 
next-to-leading-order perturbative-quantum-chromodynamics (NLO pQCD) 
calculation~\cite{NLOpQCD,Werner}.  The cutoff mass scale dependence of 
the calculation is also shown for three cases: $\mu=0.5p_T,1.0~p_T$ and 
$2.0~p_T$. The expectation for $d$+Au is calculated by scaling with the 
nuclear overlap function calculated from a Glauber model~\cite{Glauber}, 
which is expressed as $T_{d{\rm A}} = N_{\rm coll}/\sigma_{pp}^{inel}$. 
Here, 
$N_{\rm coll}$ is the number of binary nucleon-nucleon collisions and 
$\sigma_{pp}^{inel}$ is the cross section of inelastic $p$+$p$ collisions 
of 42~mb. The $p$+$p$ data points were much improved statistically 
compared to the previously published data, especially above 3~GeV/$c$, and 
the $p$+$p$ result is in good agreement with the NLO pQCD expectations. 
The $d$+Au data are higher than the expectation in $p_T<4~$GeV/$c$. Their 
$p_T$ dependence is similar to the NLO pQCD expectation, unlike the 
Au$+$Au data.


While the virtual-photon analysis suffers from low statistics at 
$p_T>5.0~$GeV/$c$, the real-photon analysis is robust at high $p_T$. The 
primary detector for the real-photon analysis is the EMCal, which 
comprises six sectors of lead-scintillator calorimeter and two sectors of 
lead-glass calorimeter.  Contamination from charged hadrons is eliminated 
by a track-matching veto in the drift chamber as well as a profile cut on 
the EMCal shower. Analysis details have been described in 
\cite{ppg060,ppg054}. The key to the method is the precise subtraction of 
the large photonic background originating from hadronic decays, about 80\% 
of which come from $\pi^0\rightarrow2\gamma$ and about 15\% from 
$\eta\rightarrow2\gamma$. Two techniques, $\pi^0$-tagging and statistical 
subtraction methods, are used to remove decay photons.


The $\pi^0$-tagging method identifies neutral pions by reconstructing 
pairs of photons in the lead-scintillator EMCal sectors that deposit more 
than 150~MeV. All pairs of photons at least 10 towers 
($\approx$0.1~radian) inside the edge of the EMCal which reconstruct to 
invariant mass $105<m_{\gamma\gamma}<165~$MeV are tagged as $\pi^0$ 
decays. The number of direct photons, $\gamma_{dir}$, is determined as

\begin{equation}
\gamma_{dir} = \gamma_{incl} - 
(1+R_{h/\pi^0})(1+\delta_{miss})\gamma_{\pi^0\rightarrow2\gamma},
\label{eq:cal_dirp_pi0}
\end{equation}

\noindent where $\gamma_{incl},\gamma_{\pi^0\rightarrow2\gamma}$ are the 
number of inclusive and $\pi^0$ decay photons, respectively, and 
$R_{h/\pi^0}$ is the ratio of other hadronic contributions to $\pi^0$ 
decay photons. $\delta_{miss}$ represents the probability that either of 
the photons from $\pi^0\rightarrow2\gamma$ misses the detector. A fast 
Monte Carlo (MC) simulation, which includes the geometric acceptance and 
EMCal response, is used to estimate $\delta_{miss}$. The input $p_T$ 
distribution of $\pi^0$ is taken from $p$+$p$ collisions~\cite{ppg024}. 
$\delta_{miss}$ is then determined as a function of $p_T$ and its 
uncertainty is evaluated as $\sim$6--8\% by varying the implemented 
simulation conditions. $R_{h/\pi^0}$ is calculated using the yield ratios 
of $\eta$ and $\omega$ to $\pi^0$ measured by PHENIX~\cite{ppg024,ppg118}.


The statistical subtraction method~\cite{ppg042,ppg049} is applied to MB 
triggered data from both the lead-scintillator and lead-glass EMCal. The 
hadron decay contribution is estimated by a hadronic cocktail simulation 
based on the observed $p_T$ spectrum of $\pi^0$; other particle spectra 
are based on the $\pi^0$ using $m_T$ scaling~\cite{ppg088}. The acceptance 
and shower merging effects are also implemented in the simulation. A 
double ratio, $R_{\gamma}$, is calculated as

\begin{equation}
R_{\gamma} = 
\left(\frac{dN_{\gamma}/dp_T}{dN_{\pi^0\rightarrow2\gamma}/dp_T}\right)^{data}/
\left(\frac{dN_{\gamma}/dp_T}{dN_{\pi^0\rightarrow2\gamma}/dp_T}\right)^{sim}.
\end{equation}

\noindent An excess due to direct photons gives $R_{\gamma}>1$, and the 
direct photon yield is determined by 
$\gamma_{dir}=(1-R_{\gamma}^{-1})\gamma_{incl}$.


\begin{figure}[tb]
\includegraphics[width=1.0\linewidth]{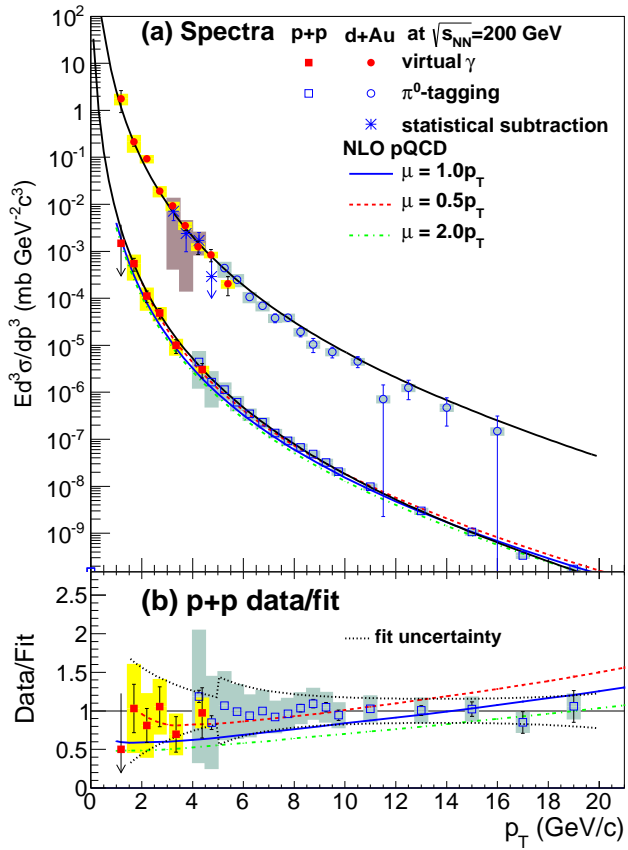}
\caption{\label{fig:spectra} (color online) (a) The invariant cross 
sections of the direct photon in $p$+$p$~\cite{ppg060,ppg136AXV} and 
$d$+Au collisions. The $p$+$p$ fit result with the empirical 
parameterization described in the text is shown as well as NLO pQCD 
calculations, and the scaled $p$+$p$ fit is compared with the $d$+Au data. 
The closed and open symbols show the results from the virtual photon and 
$\pi^0$-tagging methods, respectively. The asterisk symbols show the 
result from the statistical subtraction method for $d$+Au data, 
overlapping with the virtual photon result in $3<p_T<5~$GeV/$c$. The bars 
and bands represent the point-to-point and $p_T$-correlated uncertainties, 
respectively. (b) The $p$+$p$ data over the fit. The uncertainties of the 
fit due to both point-to-point and $p_T$-correlated uncertainties of the 
data are summed quadratically, and the sum is shown as dotted lines. The 
NLO pQCD calculations divided by the fit are also shown. 
}
\end{figure}

Figure~\ref{fig:spectra} shows the direct photon cross sections in $p$+$p$ 
and $d$+Au collisions from both virtual- and real-photon 
analyses~\cite{ppg136AXV}. The NLO pQCD calculations agree with the 
$p$+$p$ data well for a wide $p_T$ range, and show a preference for the 
choice $\mu=0.5p_T$. Unfortunately, the NLO pQCD calculation with a low 
mass cutoff scale less than $1.0~p_T$ is not available for 
$p_T<2.0~$GeV/$c$. Thus, we use an empirical parameterization, 
Eq.~\ref{eq:pp_fit}, inspired by a NLO pQCD formulation for 
$p$+$p\rightarrow\gamma X$~\cite{Werner}:

\begin{equation}
  E\frac{d^3\sigma}{dp^3} = a\cdot p_T^{-(b+c\cdot \ln x_T)} \cdot(1-x_T^2)^n,
\label{eq:pp_fit}
\end{equation}

\noindent where $a, b, c,$ and $n$ are free parameters and 
$x_T=2p_T/\sqrt{s}$.  The first factor, $p_T^{-(b+c\cdot \ln x_T)}$, is a 
power law with a logarithmic scaling correction.  The convolution of two 
PDFs in colliding protons consequently introduces the factor, 
$(1-x_T^2)^n$, which naturally leads to a drop of the cross section to 0 
at $x_T=1$.  The virtual-photon ($1.5<p_T<5~$GeV/$c$) and real-photon 
($p_T>5~$GeV/$c$) results are fit simultaneously, and the point-to-point 
uncertainty of the data is considered at fitting. The $p_T$-correlated 
uncertainty of the fit is identical with that of the data. The quadratic 
sum of these fit uncertainties is indicated as dotted lines in 
Fig.~\ref{fig:spectra}. The fit describes the data very well for the 
entire $p_T$ range. The fit parameters with uncertainty (excluding the 
$p_T$-correlated uncertainty) are 
$a$=6.6$\pm$3.3)$\times10^-3$, 
$b$=6.4$\pm$0.3, 
$c$=0.4$\pm$0.2, and 
$n$=17.6$\pm$14.9, with 
$\chi^2$/NDF=22.4/16. 
The factor of the power law, $b+c\cdot \ln x_T$, becomes 4.6--5.5 for 
$0.01<x_T<0.1$.


The $d$+Au data illustrate full consistency between the three 
aforementioned independent analyses. The independent results are in good 
agreement in the overlap region from $3.0<p_T<6.0~$GeV/$c$. The virtual 
photon analysis reaches down to 1~GeV/$c$, and the $\pi^0$-tagging method 
extends to 16~GeV/$c$. The $d$+Au data are in agreement with the binary 
collision scaled $p$+$p$ fit result across the entire $p_T$ coverage. A 
power law fit, $Ap_T^{-n}$, is performed with the $d$+Au data for 
$p_T>8~$GeV/$c$ as done for $p$+$p$
($n=7.08\pm0.09^{\rm stat}\pm0.1^{\rm syst}$)~\cite{ppg136AXV} and Au$+$Au 
($n=7.18\pm0.14^{\rm stat}\pm0.06^{\rm syst}$ for most central)~\cite{ppg139AXV}. 
The fit gives a power of $n=7.17\pm0.76^{\rm stat}\pm0.01^{\rm syst}$, consistent 
with $p$+$p$ and Au$+$Au.

\begin{figure}[tb]
\includegraphics[width=1.0\linewidth]{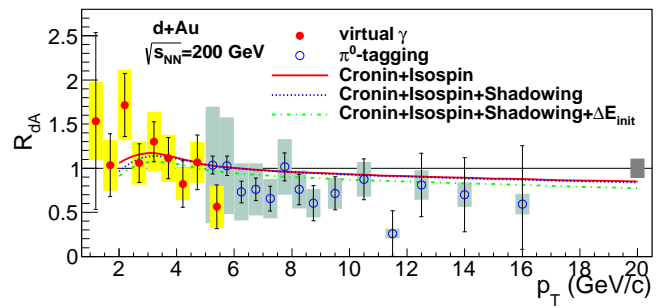}
\caption{\label{fig:dAu_ratio} (color online) 
Nuclear modification factor for $d$+Au, $R_{d{\rm A}}$, as a function of 
$p_T$. The closed and open symbols show the results from the virtual- and 
real-photon measurements, respectively. The bars and bands represent the 
point-to-point and $p_T$-correlated uncertainties, respectively. The box 
on the right shows the uncertainty of $T_{d{\rm A}}$ for $d$+Au. The 
curves indicate the theoretical calculations~\protect\cite{theo_rda} with 
different combinations of the CNM effects such as the Cronin enhancement, 
isospin effect, nuclear shadowing and initial state energy loss.
}
\end{figure}

\begin{figure}[tb]
\includegraphics[width=1.0\linewidth]{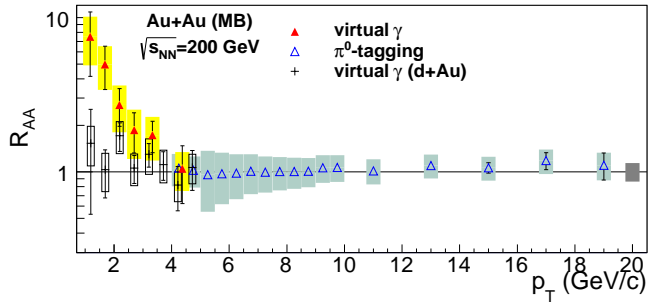}
\caption{\label{fig:AuAu_ratio} (color online) 
Nuclear modification factors for Au$+$Au (MB) and $d$+Au as a function of 
$p_T$.  The triangle symbols show results from the (closed) 
virtual~\protect\cite{ppg086} and (open) real 
photon~\protect\cite{ppg139AXV} measurements, respectively.  The bars, 
bands, and box represent the same uncertainties as in 
Fig.~\protect\ref{fig:dAu_ratio}.  The (+) symbols for $R_{d{\rm A}}$ for 
$p_T<5~$GeV/$c$ illustrates the difference in magnitude for \raa between 
Au$+$Au 
and $d$+Au collisions.
}
\end{figure}

Figure~\ref{fig:dAu_ratio} shows the nuclear modification factor for 
$d$+Au, $R_{d{\rm A}}$, calculated as the $d$+Au data divided by the 
binary-scaled $p$+$p$ fit. The point-to-point and $p_T$-correlated 
uncertainties of the $p$+$p$ fit are quadratically summed with those of 
the $d$+Au data points. The sums are shown as bars and bands, 
respectively. The uncertainty of $T_{d{\rm A}}$ for $d$+Au is indicated by 
the box located at the right in Fig.~\ref{fig:dAu_ratio}. $R_{d{\rm A}}$ 
is consistent with unity within the reported uncertainty. The theory 
calculations~\cite{theo_rda} with different combinations of standard CNM 
effects are shown with the data. The solid curve is the simplest one 
including only the Cronin enhancement and isospin effect. The nuclear 
shadowing with the EKS98 parameterization~\cite{EKS98} of the nPDFs is 
additionally considered for the dotted and dash-dotted curves, and the 
initial state energy loss is included for the dash-dotted curve. The data 
are consistent within uncertainties with the theoretical calculations, but 
do not have a sufficient precision to resolve the considered initial state 
nuclear effects. The data do however rule out much larger effects beyond 
these standard range predictions.  In contrast, Fig.~\ref{fig:AuAu_ratio} 
shows that for $R_{\rm AA}$ in Au$+$Au collisions, there is a much larger 
enhancement of the direct photon production below 2.0~GeV/$c$.  The 
magnitude of the enhancement in Au$+$Au with $R_{\rm AA}>7$ is much higher 
than observed in $d$+Au, indicating that there is a significant medium 
effect on direct photon production.

In conclusion, direct photons in $1<p_T<16~$GeV/$c$ have been measured for 
$d$+Au collisions via three independent analyses, the virtual photon, 
$\pi^0$-tagging and statistical subtraction methods. The results from 
these analyses agree in the overlap $p_T$ region. The $p$+$p$ spectrum has 
also been improved statistically by the 2006 data. The improved $p$+$p$ 
data are parameterized by a pQCD inspired fit function. The fit describes 
the data very well for the entire $p_T$ region. $R_{d{\rm A}}$ is 
consistent with unity.  The data fully support the theoretical 
calculations with the standard CNM effects for a wide $p_T$ range. 
$R_{\rm AA}$ shows a much larger enhancement below 2.0~GeV/$c$ compared to 
the $d$+Au data, indicating the existence of a medium effect as an 
additional source of direct photons.




We thank the staff of the Collider-Accelerator and Physics
Departments at Brookhaven National Laboratory and the staff of
the other PHENIX participating institutions for their vital
contributions.  We acknowledge support from the 
Office of Nuclear Physics in the
Office of Science of the Department of Energy,
the National Science Foundation, 
a sponsored research grant from Renaissance Technologies LLC, 
Abilene Christian University Research Council, 
Research Foundation of SUNY, 
and Dean of the College of Arts and Sciences, Vanderbilt University 
(U.S.A),
Ministry of Education, Culture, Sports, Science, and Technology
and the Japan Society for the Promotion of Science (Japan),
Conselho Nacional de Desenvolvimento Cient\'{\i}fico e
Tecnol{\'o}gico and Funda\c c{\~a}o de Amparo {\`a} Pesquisa do
Estado de S{\~a}o Paulo (Brazil),
Natural Science Foundation of China (P.~R.~China),
Ministry of Education, Youth and Sports (Czech Republic),
Centre National de la Recherche Scientifique, Commissariat
{\`a} l'{\'E}nergie Atomique, and Institut National de Physique
Nucl{\'e}aire et de Physique des Particules (France),
Bundesministerium f\"ur Bildung und Forschung, Deutscher
Akademischer Austausch Dienst, and Alexander von Humboldt Stiftung (Germany),
Hungarian National Science Fund, OTKA (Hungary), 
Department of Atomic Energy and Department of Science and Technology (India),
Israel Science Foundation (Israel), 
National Research Foundation and WCU program of the 
Ministry Education Science and Technology (Korea),
Ministry of Education and Science, Russian Academy of Sciences,
Federal Agency of Atomic Energy (Russia),
VR and Wallenberg Foundation (Sweden), 
the U.S. Civilian Research and Development Foundation for the
Independent States of the Former Soviet Union, 
the US-Hungarian Fulbright Foundation for Educational Exchange,
and the US-Israel Binational Science Foundation.


\def\NIM{Nucl. Instrum. Methods}
\def\NIMA{Nucl. Instrum. Methods~A}
\def\NPA{Nucl. Phys.~A}
\def\NPB{Nucl. Phys.~B}
\def\EPJ{Eur. Phys.~J. C}
\def\RMP{Rev.~Mod.~Phys. }
\def\SOV{Sov. J. Nucl. Phys. }
\def\PLB{Phys.~Lett.~B}
\def\PR{Phys.~Rpts.}
\def\PRL{Phys. Rev. Lett.}
\def\PRD{Phys. Rev.~D}
\def\PRC{Phys. Rev.~C}
\def\ZPC{Z.~Phys.~C}

\bibliographystyle{apsrev}


\end{document}